\documentclass[twocolumn,showpacs,preprintnumbers,amsmath,amssymb,prb]{revtex4}

\usepackage{graphicx}
\usepackage{dcolumn}
\usepackage{bm}
\newcommand{\dir}{Figs}

\begin{document}


\title{Molecular recognition in a lattice model: An enumeration study}

\author{Thorsten Bogner, Andreas Degenhard, Friederike Schmid}
\affiliation{
Condensed Matter Theory Group, Fakult\"at f\"ur Physik, Universit\"at Bielefeld
}
\date{\today}

\begin{abstract}
We investigate the mechanisms underlying selective molecular recognition of 
single heteropolymers at chemically structured planar surfaces. To this end, 
we study systems with two-letter (HP) lattice heteropolymers by exact
enumeration techniques. Selectivity for a particular surface is defined by
an adsorption energy criterium. We analyze the distributions of selective
sequences and the role of mutations. A particularly important factor for
molecular recognition is the small-scale structure on the polymers.
\end{abstract}

\maketitle
Selective molecular recognition governs many biological processes
such as DNA-protein binding~\cite{zakrzewska_2003a} or cell-mediated
recognition~\cite{alberts_1994a}. Biotechnological applications
range from the development of biosensoric materials~\cite{nakata_2004a}
to cell-specific drug-targeting~\cite{christi_2004a}. The specificity
in these processes results from the interplay of a few unspecific 
interactions (van der Waals forces, electrostatic forces, hydrogen
bonds, and the hydrophobic force)~\cite{israelachvili_1991a}
and a heterogeneous composition of the polymer chain. Selectivity 
is a genuinely cooperative effect. The question how it emerges in
a complex system is therefore very interesting from the point 
of view of statistical physics, and the study of idealized models 
can provide insight into general 
principles~\cite{janin_1996a,muthukumar_1999a,chakraborty_2001a,lee_2003a}.

Previous theoretical studies have mostly considered heteropolymer
adsorption in either regular~\cite{muthukumar_1995a} or 
random~\cite{janin_1996a,chakraborty_2001a,polotsky_2004a,polotsky_2004b}
systems. The interplay of cluster sizes on random heteropolymers and random 
surfaces and its influence on the adsorption thermodynamics and kinetics 
was studied analytically and with computer 
simulations~\cite{chakraborty_2001a,polotsky_2004b,golumbfskie_1999a}. 
Concepts from the
statistical physics of spin glasses were used to study the adsorption of
polymers on a ``native'' surface compared with that on an arbitrary
random surface~\cite{janin_1996a,srebnik_1996a,chakraborty_2001a}.

In the present paper, we focus on a different question: We investigate
mechanisms by which specific heteropolymers {\em distinguish}
between given surfaces. To this end, we adopt an approach which has proven 
highly rewarding in the context of the closely related problem of protein 
folding~\cite{dill_1985a, lau_1989a, li_1996a, li_1997a}:
We enumerate exactly all compact polymer conformations within a lattice model.
The protein is described as a heteropolymer chain consisting of two types 
of monomers, hydrophobic (H) and polar (P), which occupy each one site on the
lattice. Sites surrounding the polymer are assumed to contain solvent. 
The protein is exposed to an impenetrable flat surface covered with 
sites of either type H or type P, which form a particular surface 
pattern. It may adsorb there and change its conformation during the 
adsorption process. However, we require that both the free and the adsorbed 
chain are compactly folded in a given shape
(cubic or rectangular)~\cite{li_1996a,li_1997a}.
Nearest neighbor particles interact with fixed, type dependent interaction 
energies.  Surface sites H and P are considered to be equivalent to monomer 
sites H and P. The total energy is then given by:
\begin{equation}
  E_{\rm tot} \;=\; \sum\limits_{<i,j>}\sum\limits_{\alpha, \beta}
                   \tau_i^{\alpha}\tau_j^{\beta}E_{\alpha \beta}
 \label{etot}
\end{equation}
Here the sum $<i,j>$ runs over nearest neighbor pairs, the sums
$\alpha$ and $\beta$ run over the types hydrophobic (H), polar (P), 
or solvent (S), and $\tau_i^{\gamma}$ is an occupation number 
which takes the value one if the site $i$ is occupied with type 
$\gamma$, and zero otherwise. For compact chains with a fixed 
sequence, the energy spectrum as defined by Eq. (\ref{etot}) is (except
for a fixed offset) fully characterized by only two parameters: One which
describes the relative incompatibility of $H$ and $P$ inside the globule,
$V=2E_{HP}-E_{HH}-E_{PP}$, and one which accounts for the difference
between the affinities of $H$ and $P$ to the solvent,
$W=2(E_{HS}-E_{PS})+(E_{PP}-E_{HH})$. Since one of these parameters sets
the energy scale, the model has only one dimensionless free parameter,
$V/W$. Motivated by Ref.~\cite{li_1996a}, where $V/W=0.13$,
we chose $V/W=0.1$.


We consider two-dimensional and three-dimensional systems with
system sizes up to $6 \times 6$ (in 2D) and $3 \times 3 \times 3$
(in 3D), respectively. For each system, a set of sequences was picked 
randomly (uncorrelated monomers, equal probability for H and P). 
For each sequence, we then evaluated the energies for all possible compact 
chain conformations in contact with all possible surfaces.
This allowed us to determine exactly the ground-state adsorption
energy on every surface. We call a sequence {\em selective}, if there
exists one unique surface with highest adsorption energy, {\em i.e.},
if the difference
\begin{equation}
  E_{\rm gap} \;=\; E_{\rm ad}^{1st} \,-\, E_{\rm ad}^{2nd} \;.
 \label{energy_gap}
\end{equation}
between the adsorption energies on the two most favorable surfaces
is nonzero. The lowest-energy {\em structure} of the chain on its 
favorite surface (the ``selected'' surface) is not necessarily unique.

We note that this selectivity criterion is a ``zero-temperature''
criterion. Entropic contributions to the adsorption free energy
are not accounted for. Furthermore, we disregard dynamic and kinetic 
factors~\cite{sommer_2002a}, which presumably also play a role in 
molecular recognition processes. 



\begin{figure}[b]
  \includegraphics[width=0.45\textwidth,height=0.4\textwidth]{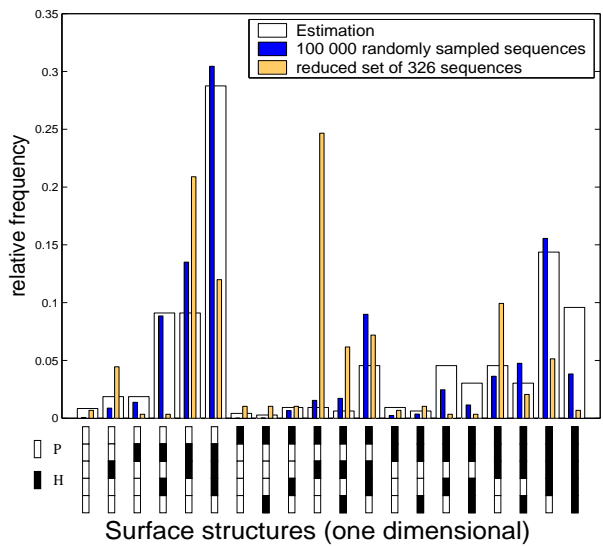}
\caption{Relative frequency of sequences selective for different
         surfaces on the $5\times5$ lattice. The black bars show
         the result for a random sample, the gray bars for a 
         sample based on a ``master sequence''. Also shown are
         the values obtained with a least-square fit to 
         Eq.~(\ref{estimate}) (see text for explanation).}
\label{fig:surface_distrib}
\end{figure}

In all systems, more than $90 \%$ of all sequences were selective. 
The distribution on the different surfaces was highly inhomogeneous,
see Fig.~\ref{fig:surface_distrib}.
A closer inspection reveals that two main factors contribute to 
the frequency with which sequences select a particular surface pattern:
A high number of hydrophobic sites inside the pattern is 
beneficial, whereas hydrophobic sites at the border are unfavorable.
This is due to the fact that bound proteins prefering the latter surface
patterns must have hydrophobic monomers at the edges.
The resulting unfavorable contacts to the solvent have to be compensated to
achieve an energetic minimum. This reduces the number of suited sequences. 
The frequency distribution could be fitted remarkably well
by the simple formula
\begin{equation}
\label{estimate}
N=\frac{An_{core}^{m}+B} {n_{border}+1},
\end{equation}
where $n_{core}$ denotes the number of hydrophobic core sites, 
$n_{border}$ the number of hydrophobic border sites, $m$ the
total number of core sites, and $A$ and $B$ are fit parameters.
For the $5\times 5$ system, such a fit is illustrated in 
Fig.~\ref{fig:surface_distrib}. The fitting is also successful for
other systems, even for the 3D case, if one identifies sites at the
corner of the surface with border sites. The functional form of
Eq.~(\ref{estimate}) was guessed empirically, with no
underlying theory, and should not be over-interpreted. 
Nevertheless, we can conclude that the relative frequency 
of surface patterns is mostly determined by a few, 
unspecific surface characteristics.

The previous analysis raises the question how sequences which are
selective for different surfaces differ from one another, or, conversely, 
which features sequences belonging to the same surface have in common.
We have used different approaches to address this problem.

The first approach was motivated by the biological principle of mutation.
A similarity measure between two chain sequences can be defined by 
counting the minimum number of point mutations required to construct
one sequence, starting from the other. For our two-letter sequences,
this is quantified using the {\em Hamming distance}
\begin{equation}
  d(s,s') \;:=\; \frac{1}{2} \sum_{i} \left| s_i - s'_i \right| \;,
 \label{hamm-def}
\end{equation}
between sequences $s$ and $s'$. The sum $i$ runs over all monomers
along the chain, and the variables $s_i$, $s_i'$ are taken to be $s_i,s_i'=1$ 
if the $i$th monomer of the sequence $s$ is hydrophobic, and $s_i,s_i'=-1$ otherwise.
Two sequences that have a Hamming distance of $n$ are thus separated by 
$n$ point mutations. Since sequences can be read in both directions,
Eq.~(\ref{hamm-def}) usually yields two values for a pair of sequences. 
We have always used the smaller one.
\begin{figure}[t]
\includegraphics[width=0.45 \textwidth]{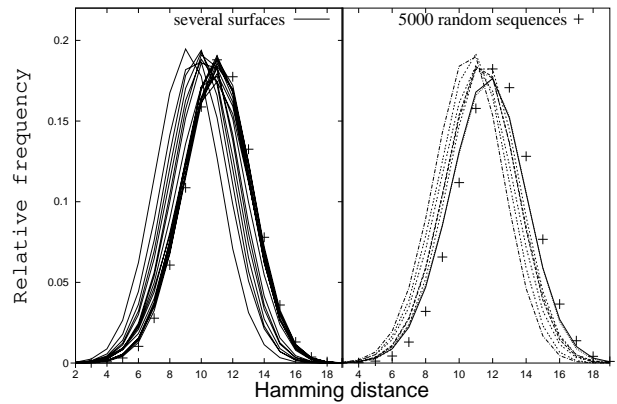}
\caption{Histograms showing the distribution of Hamming distances for
  the $5\times5$ (left) and the $3\times3\times3$ (right) lattice
  in set of sequences belonging to several surfaces (lines).
  Also shown for comparison are the results for a set of 5000
  random sequences (crosses).}
\label{fig:hamm-dist}
\end{figure}

Based on this definition, we can now study whether sequences 
belonging to the same surface are ``close'' in sequence space. 
Examples of distributions of Hamming distances for different
surfaces are shown in Fig.~\ref{fig:hamm-dist}. The distributions
for different surfaces, and even for different system sizes, are 
very similar. The number of mutations with the highest occurrence is nearly 
half the total number of monomers in the polymer chain. Moreover,
the distribution is not very different from that of a totally random
set of sequences, which is also shown in Fig.~\ref{fig:hamm-dist} for comparison. 
Hence we conclude that the sequences selective for a particular surface 
are widely distributed over the sequence space, and that proximity in 
sequence space is not a relevant factor for molecular recognition.

This result has interesting practical consequences. An important 
issue for many cell-surface recognition processes is the question 
how efficiently nature distinguishes between different 
surfaces~\cite{davis_2003a}, {\em i.e.}, how many
mutations are required to change a polymer sequence that is selective
for a particular surface to make it selective to another surface or
a whole class of different surfaces. In our model, the observation that 
sequences selective to the {\em same} surface appear to be widely
spread in sequence space suggests that one might find sequences 
which are selective to very {\em different} surfaces at close 
vicinity in sequence space.

In order to test this idea,
we have attempted to compute subsets of sequences, which are close
in sequence space and nevertheless ``recognize'' all surfaces, 
{\em i.e.}, which contain at least one selective sequence 
for each surface. Such sets were constructed following a two-step 
procedure. First, we identified a {\em center} or {\em master} sequence,
which was a suitable initial point for the mutation process. This
was done mainly by trial and error, starting from the sequences belonging
to the least favorable surfaces.  Second, we evaluated the number of 
mutations necessary to provide a subset of sequences recognizing all 
surfaces. This analysis was carried out for different two-dimensional
systems. The results are shown in table~\ref{tab:radii}. In spite of
the exponential growth in the number of possible polymer chain
conformations and possible sequence realizations, the number of necessary
mutations $r$ in table~\ref{tab:radii} increases only slightly with
the surface size. The distribution of the sequences on the surfaces
is shown for one of these reduced subsets in Fig.~\ref{fig:surface_distrib},
and can be compared with the full distribution. 
The general features are comparable.

\begin{table}[b]
  \vspace*{-0.4cm}
  \caption{\label{tab:radii}Number of mutations $r$ necessary to generate
    a subset of sequences which recognize all surfaces, together
    with the corresponding subset size for various lattice sizes.
    In the case of rectangular folding ($5\times4$ and $6\times5$),
    the largest side forms the interface to the surface.}
  \begin{ruledtabular}
    \begin{tabular}{lccr}
      Lattice&surface size&r&size of set
      \\\hline
      $5\times4$   &     5 & 2 &209\\
      $5\times5$   &     5 & 2 &326\\
      $6\times5$   &     6 & 3 &466\\
      $6\times6$   &     6 & 4 &7807\\
    \end{tabular}
  \end{ruledtabular}
\end{table}

We note that the values $r$ for the minimum number of mutations required
to {\em recognize} all surfaces, as given in table~\ref{tab:radii}, are 
upper limits and can possibly be reduced further with more efficient master 
sequences. Even so, $r$ is in some cases smaller than the minimum number
of mutations necessary to {\em generate} all surfaces (starting from
a common master surface). Hence only a few point mutations can alter 
the adsorption characteristics profoundly. This result matches with
experimental results obtained from binding force measurements on 
antibodies~\cite{ros_1998a}.
Experimentally, it was observed that the wild-type antibody and a mutant
in which an amino acid at one position in the chain has been exchanged
differ in the measured affinity by roughly one order of magnitude.

\begin{figure}[t]
  \includegraphics[width=0.3 \textwidth, angle=270]{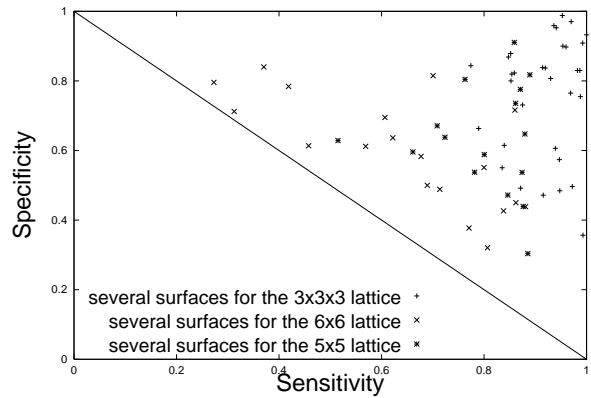}
\caption{Performance of the fully-connected two-layer
  perceptron trained for several surface structures on the $5\times5$ and
  $3\times3\times3$ lattice displayed in a sensitivity (true positive) versus
  specificity (true negative) plot. The diagonal line represents results with
  a $50\%$ correct classification rate corresponding to random guessing.
  For the $6\times6$ system the results were obtained by a
  fully-connected three-layer perceptron with 16 hidden units.
  In all cases the data have been transformed to Fourier space and the
  perceptron was optimized via a backpropagation algorithm, see
  Ref.~\cite{bishop_1995a}}\label{fig:net_result}
\end{figure}

We return to the problem of determining common features of sequences 
which are selective for the same surface.
To clarify the question whether there exist any such features,
we have applied an artificial neural network (ANN).
After training the ANN with a set, composed equally from
selective as well as non selective sequences for a given surface,
the performance of the ANN was tested with a second, disjoint set. 
This analysis was performed for all surfaces with at least 100 
selective sequences.
The results of the testing, Fig.~\ref{fig:net_result},
show that there do exist relevant features for the recognition
process that can be learned by the ANN.

The next question is: What does the ANN learn? In the case of a two-layer
perceptron, the answer is relatively simple~\cite{bishop_1995a}:
The ANN classifies by dividing the sequence space of dimension $N$ into 
two parts by a $N-1$ dimensional hyper-plane. 
The fact that this classification is successful suggests that insight 
might be gained
by a more general characterization than the mere mutual (Hamming) distances.
In order to achieve this we applied the ``Principal Component Analysis''
(PCA)~\cite{jolliffe_1986a}. In this approach, the data, i.e. in the present
paper the discrete Fourier transform of the sequences, are treated as a random
vector. In general the modes are correlated, in particular if common features
within a set of sequences exist. This is characterized by the
variances and covariances given in the covariance matrix.
Diagonalization of this matrix yields a description by uncorrelated
components, the eigenvectors. The eigenvalues are a measure for
the squared variances of these components. Low eigenvalues correspond to
characteristic components within the set.


\begin{figure}[t]
\includegraphics[width=0.39\textwidth]{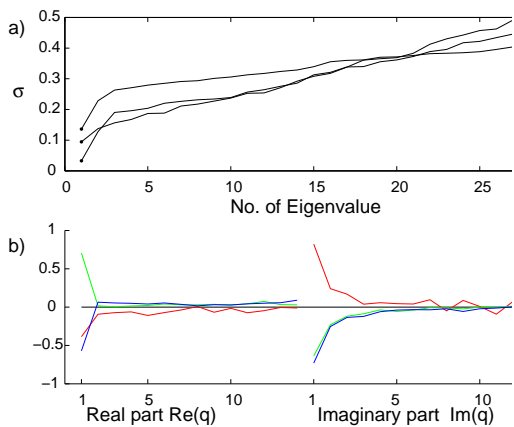}
\caption{(a) Square roots of the eigenvalues of the covariance matrix
  and (b) coordinates in Fourier space ($q$-space) of the eigenvector
  corresponding to the smallest variance (circles in a)
  for three surface patterns of the $3\times3\times3$ system.}
\label{fig:eigenvalues}
\end{figure}

\begin{figure}[b]
\includegraphics[width=0.35\textwidth]{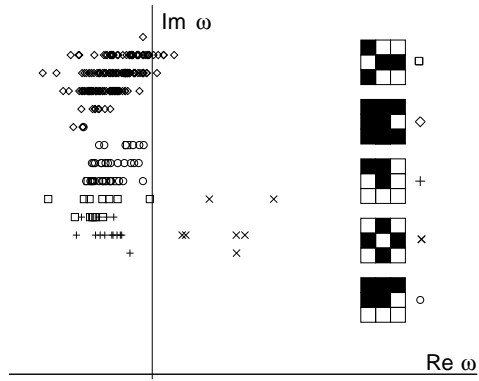}
\caption{Projection of sequences on the highest frequency ($\omega$)
         plane for the $3\times 3\times 3$ lattice and various surface
         structures. Some of the sets are completely separated in this plane.}
\label{fig:proj_plane1}
\end{figure}

We have carried out PCAs for various surfaces in the $5 \times 5$-, the
$6 \times 6$-, and the $3 \times 3 \times 3$ system. The results revealed
an unexpected common  feature: For all surfaces in the $5 \times 5$- and the
$3 \times 3 \times 3$ system, two components turned out to be especially
meaningful, namely almost exactly the highest frequency modes (real and
imaginary part). The corresponding variances were considerably smaller
than those of all other components, see  Fig.~\ref{fig:eigenvalues}. 
In the $6 \times 6$ system, the result was not as simple, yet the
high-frequency components were still among the significant components.

These results can be visualized by projecting the sequence space onto
the highest-frequency plane.
Fig.~\ref{fig:proj_plane1} illustrates for the $3\times3\times3$-system
that sets of sequences belonging to different surfaces often occupy
different regions in this plane. 

To summarize, we have studied the recognition of chemically structured
surfaces by single polymer chains comprising hydrophilic and hydrophobic
monomer units. Starting from already folded conformations, we investigated
distributions of selective sequences and the role of point mutations. We found
that sequences recognizing the same surface are widely distributed in
sequence space, {\em i.e.}, they are separated by many mutations. 
Conversely, it was in many cases possible to construct a subset of sequences
which recognize all surfaces and nevertheless differ from one another
by only a few mutations. Despite their wide distribution, sequences
recognizing the same surface have features in common, which can be
learned by a neural network. One factor which turned out to be particularly
important in this recognition process is the local, small-scale structure 
on the polymers.

We thank Alexey Polotsky for useful discussions and the german science
foundation (DFG) for partial support.

\bibliography{paper}

\end{document}